\documentclass[preprint,12pt]{elsarticle}

\usepackage{amsmath}
\usepackage{amssymb}
\usepackage{amsthm}

\usepackage{xcolor}

\newcommand{\dps}{\displaystyle}

\def\argS{(k_1,\varepsilon_1;\ldots ; k_S,\varepsilon_S)}

\newcommand{\be}{\begin{equation}} 
\newcommand{\ee}{\end{equation}} 
\newcommand{\bea}{\begin{eqnarray}} 
\newcommand{\eea}{\end{eqnarray}}

\def\slash#1{#1\!\!\!\raise.15ex\hbox {/}}
\newcommand{\slD}{\,\raise.15ex\hbox{$/$}\kern-.27em\hbox{$\!\!\!D$}}
\newcommand{\slpartial}{\raise.15ex\hbox{$/$}\kern-.57em\hbox{$\partial$}}

\def\2int{\int_{0}^{1}  \!\!\! du \! \int_{0}^{1}  \!\!\! dv}

\def\e{\mbox{e}}

\def\Z{{\mathchoice {\hbox{$\sf\textstyle Z\kern-0.4em Z$}}
{\hbox{$\sf\textstyle Z\kern-0.4em Z$}}
{\hbox{$\sf\scriptstyle Z\kern-0.3em Z$}}
{\hbox{$\sf\scriptscriptstyle Z\kern-0.2em Z$}}}}

\def\no{\noindent}
\def\non{\nonumber\\}

\def\half{\frac{1}{2}}

\def\fourth{\frac{1}{4}}

\def\mn{{\mu\nu}}

\def\e{\,{\rm e}}

\def\b0{{\bf 0}}

\def\bear{\begin{eqnarray}}
\def\ear{\end{eqnarray}\noindent}
\def\bec{\blue\begin{equation}}
\def\eec{\end{equation}\black\noindent}
\def\bearc{\blue\begin{eqnarray}}
\def\earc{\end{eqnarray}\black\noindent}
\def\benn{\begin{enumerate}}
\def\enn{\end{enumerate}}

\def\totint{\int_{-\infty}^{\infty}}

\def\intdp3{\int\frac{d^3p}{(2\pi)^3}}
\def\intdp4{\int\frac{d^4p}{(2\pi)^4}}

\def\kinb{\frac{\dot x^2}{4}}

\def\4piTD{{(4\pi T)}^{-{D\over 2}}}
\def\4piT4{{(4\pi T)}^{-2}}
\def\Tint{{\dps\int_{0}^{\infty}}\frac{dT}{T}}

\newcommand{\slG}{{{\dot G}\!\!\!\! \raise.15ex\hbox {/}}}

\def\ddtau{{d\over d\tau}}

\def\dps{\displaystyle}

\def\Nint{\prod_{i=1}^N\int_0^Td\tau_i}

\begin{document}

\begin{frontmatter}



\title{$N$-photon amplitudes in a plane-wave background}

\author[label1]{James P. Edwards}
 \affiliation[label1]{organization={Instituto de F\'isica y Matem\'aticas,Universidad Michoacana de San Nicol\'as de Hidalgo},
             addressline={Edificio C-3, Apdo. Postal 2-82},
             city={Morelia},
             postcode={58040},
             state={Michoac\'an},
             country={Mexico}}


\author[label1]{Christian Schubert}


\begin{abstract}
We use the worldline formalism to derive master formulas for the one-loop $N$ - photon amplitudes in a plane-wave
background, for both scalar and spinor QED. This generalises previous work by Ilderton and Torgrimsson
for the vacuum polarisation case, although with some change in methodology since, instead of evaluating the path integral
on the semi-classical trajectory, we use the special kinematics of the plane-wave background to uncover the
crypto-gaussian character of this type of worldline path integral. 
\end{abstract}
%
%
%
%
%
%

\end{frontmatter}


\section{Introduction}
\label{intro}

In strong-field QED, there are two external field configurations that play a special r\^ole: the constant field, and the plane-wave one.
Both are not only important from the physics point of view, but also special mathematically, since they allow for an exact solution
of the Dirac equation in the field, which makes it possible to perform non-perturbative calculations in such fields
(see, e.g., \cite{frgish-book,ditgie-book}). 
Nevertheless, beyond the simplest special cases such calculations tend to be extremely lengthy and tedious 
\cite{nikrit1,nikrit2,baibre,adler71,tsai,bks1,bks2}.
For example, the one-loop QED vertex in a plane-wave field has been calculated only very recently \cite{pialop}. 

For the constant-field case there exists an alternative approach, based on Feynman's worldline path integral formulation of
QED \cite{feynman:pr80,feynman:pr84}
and concepts originally borrowed from string theory \cite{berkosNPB,strassler1}, that has been shown to offer various technical
advantages for closed-loop photonic processes \cite{5,shaisultanov,17,18} and recently also for amplitudes involving open scalar \cite{110} and
fermion \cite {113} lines -- for reviews of this formalism see \cite{41, Edwards:2019eby}.

The plane-wave case has attracted much attention in recent years because of its relevance for laser physics
\cite{rituslaser,ehkrka,pmhk}. 
However, the application of the worldline formalism to this case has turned out to be 
less straightforward. A calculation of the scalar and spinor QED vacuum polarisations along these lines was achieved
by A. Ilderton and G. Torgrimsson \cite{ildtor}, but it is not obvious how to extend their approach to the general $N$-photon
amplitudes. Here we will use a slightly different approach, based on a direct rewriting of the worldline path integral
as a gaussian one, to construct compact master formulas for the scalar and spinor QED $N$-photon amplitudes 
in a plane-wave background. 

We start in the following section with a short summary of the worldline representation of the $N$-photon amplitudes in 
vacuum (for details see \cite{41}). 
The following two sections are devoted to the derivation of master formulas for the $N$-photon amplitudes 
in a plane-wave background, first for scalar and then for spinor QED. As a check, in section \ref{N2} we work out the
$N=2$ cases and recover the results of \cite{ildtor}. In the final section we shortly summarize our results and point
out possible generalisations. 

\section{$N$-photon amplitudes in the worldline formalism}

The starting point for the calculation of the scalar QED $N$-photon amplitudes in the worldline formalism 
is Feynman's \cite{feynman:pr80} worldline representation of the
one-loop effective action $\Gamma_{\rm scal}[A]$:
\bear
\Gamma_{\rm scal}[A] &=& \int_0^{\infty} \frac{dT}{T}
\ \e^{-m^2T}\int_{x(T)=x(0)} Dx
\, \e^{-\int_0^T d\tau\bigl(\kinb + ie\,\dot x\cdot A(x) \bigr)}
\label{1-scalEA}
\ear
The path integral runs over all closed trajectories in spacetime obeying the 
periodicity condition $x(T)=x(0)$ in proper time. 
The $N$-photon amplitudes are obtained from this by expanding the ``interaction exponential''
%
%
%
and Fourier transformation, which leads to the ``vertex operator representation'' of the $N$-photon amplitude:
\bear
\Gamma_{\rm scal}(k_1,\varepsilon_1;\ldots ; k_N,\varepsilon_N) &=&
(-ie)^N 
\int_0^{\infty}
\frac{dT}{T}\,
\e^{-m^2T}
\int_{x(0)=x(T)}
Dx\,
\e^{-\int_0^T d\tau
\kinb}
\nonumber\\
&& \times
 V^{\gamma}_{\rm scal}[k_1,\varepsilon_1]\cdots V^{\gamma}_{\rm scal}[k_N,\varepsilon_N]
\label{Nphotonvertop}
\ear
Here each photon is represented by the following {\sl photon vertex operator}, integrated along the trajectory:
\bear
V^{\gamma}_{\rm scal}[k,\varepsilon] &\equiv &\int_0^T d\tau \,\varepsilon\cdot\dot x(\tau)\,{\e}^{ik\cdot x(\tau)}
\label{defphotonvertop}
\ear\no
%
After a formal exponentiation 
$
\varepsilon_i\cdot
\dot x_i\,\e^{ik_i\cdot x_i}
=
\e^{\varepsilon_i\cdot\dot x_i
+ik_i\cdot x_i}
\mid_{\varepsilon_i}
\label{formexp}
$
the path integral can be done by gaussian integration using the basic correlator
\bear
\langle x^\mu(\tau)x^\nu(\tau') \rangle = - G(\tau,\tau')\delta^\mn, \quad
G(\tau,\tau') \equiv \mid \tau-\tau'\mid 
-\frac{(\tau-\tau')^2}{T} 
\, .
\label{defG}
\ear
\no
This results in the following ``Bern-Kosower representation'' of the $N$-photon amplitude \cite{polyakovbook,berkosNPB,strassler1},
\begin{eqnarray}
\Gamma_{\rm scal}(k_1,\varepsilon_1;\ldots;k_N,\varepsilon_N)
&=&
{(-ie)}^N
{(2\pi )}^D\delta (\sum k_i)
{\dps\int_{0}^{\infty}}\frac{dT}{T}
{(4\pi T)}^{-\frac{D}{2}}
e^{-m^2T}
\nonumber\\
&& \hspace{-150pt}
\times
\prod_{i=1}^N \int_0^T 
d\tau_i
\exp\biggl\lbrace\sum_{i,j=1}^N 
\Bigl\lbrack  \half G_{ij} k_i\cdot k_j
-i\dot G_{ij}\varepsilon_i\cdot k_j
+\half\ddot G_{ij}\varepsilon_i\cdot\varepsilon_j
\Bigr\rbrack\biggr\rbrace
\Bigl\vert_{\varepsilon_1\varepsilon_2\ldots \varepsilon_N}
\label{scalarqedmaster}
\end{eqnarray}
\no
Here a `dot' denotes a derivative acting on the first variable,
\begin{eqnarray}
\dot G(\tau,\tau') &=& {\rm sign}(\tau - \tau')
- 2 \frac{(\tau - \tau')}{T}, \quad
\ddot G(\tau,\tau')
= 2 {\delta}(\tau - \tau')
- \frac{2}{T}
\label{GdGdd}
\end{eqnarray}
\noindent
and we abbreviate $G_{ij}\equiv G(\tau_i,\tau_j)$ etc.
The factor ${(4\pi T)}^{-\frac{D}{2}}$
represents the free Gaussian path integral
determinant factor, and the 
${(2\pi )}^D\delta (\sum k_i)$ factor is produced by the integration over the zero mode $x_0^{\mu}\equiv \frac{1}{T}\int_0^Td\tau\, x^\mu(\tau)$
of the path integral.
The exponential must still be expanded and only the terms be retained that contain
each polarisation vector $\varepsilon_i$ linearly:
\bear 
\exp\bigl\lbrace 
\cdot
\bigr\rbrace 
\bigl\vert_{\varepsilon_1\varepsilon_2\ldots \varepsilon_N}
 \quad\equiv
 {(-i)}^N P_N(\dot G_{ij},\ddot G_{ij})
 \exp\biggl[\half \sum_{i,j=1}^N G_{ij}k_i\cdot
k_j \biggr] 
\label{defPN}
\ear
with certain polynomials $P_N$. 

For spinor QED, a generalization of \eqref{1-scalEA} suitable for analytical calculations is given by the Feynman-Fradkin representation
\cite{feynman:pr84,fradkin}
\bear
\Gamma_{\rm spin}[A]
&=&
-\half \Tint
\int Dx \int D\psi \,
e^{-\int_0^T d\tau\bigl(
\fourth \dot q^2
+ \half \psi\cdot \dot\psi
+\half ie\, \dot x \cdot A(x) 
-ie \psi^{\mu}F_{\mn}(x)\psi^{\nu}\bigr)}
\, .
\non
\label{spinorQEDgrass}
\ear\no
Here $\psi^{\mu}(\tau)$ is a Lorentz vector whose components are Grassmann functions,  
$\lbrace\psi^{\mu}(\tau),\psi^{\nu}(\tau')\rbrace = 0$, and the path integral 
$\int D \psi$ has to be taken over antiperiodic such functions, $\psi^{\mu}(T) = - \psi^{\mu}(0)$.  
Note that it is already gaussian as it stands. 

Applying the same procedure as for the scalar case above, one obtains the following generalization 
of \eqref{Nphotonvertop} to the spinor QED case:
\bear
\Gamma_{\rm spin}(k_1,\varepsilon_1;\ldots ; k_N,\varepsilon_N) &=&
-\half(-ie)^N 
\int_0^{\infty}
\frac{dT}{T}\,
\e^{-m^2T}
\int_{x(0)=x(T)}
Dx\,
e^{-\int_0^T d\tau
\kinb}
\nonumber\\
&&\hspace{-50pt} \times
\int_{A} D\psi
\,
\e^{
-\int_0^Td\tau
\half \psi\cdot \dot\psi 
}
 V^{\gamma}_{\rm spin}[k_1,\varepsilon_1]\cdots V^{\gamma}_{\rm spin}[k_N,\varepsilon_N] \, .
\label{DNpointspin}
\ear
The photon vertex operator for spinor QED $V^{\gamma}_{\rm spin}$ differs from the scalar one 
\eqref{defphotonvertop} by a second term representing the interaction of the fermion spin with
the photon, 
\bear
 V^{\gamma}_{\rm spin}[k,\varepsilon] &\equiv &  \int_0^Td\tau\, \Bigl[ \varepsilon\cdot \dot x(\tau)
 -i \psi(\tau)\cdot f \cdot \psi(\tau)\Bigr] \,\e^{ik\cdot x(\tau)}
\label{defVspin}
\ear
with $f^\mn \equiv k^\mu\varepsilon^\nu - \varepsilon^\mu k^\nu$ the photon field-strength tensor. 
Thus the $N$-photon amplitude is naturally obtained in terms of a spin-orbit decomposition
\bear
\Gamma_{{\rm spin},N} = \sum_{S=0}^N \Gamma_{NS}\, ,\quad 
\Gamma_{NS} = \sum_{\lbrace i_1i_2\ldots i_S\rbrace} 
\Gamma_{NS}^{\lbrace i_1i_2\ldots i_S\rbrace}\, , 
\label{decompGammaSO}
\ear
where $S$ denotes the number of spin interactions, and the sum $ \sum_{\lbrace i_1i_2\ldots i_S\rbrace}$ 
runs over all choices of $S$ out of the $N$ photons as the ones assigned to those interactions. 
It is then straightforward to arrive at the
following master formula for $\Gamma_{NS}^{\lbrace i_1i_2\ldots i_S\rbrace}$:
\begin{align}
\Gamma_{NS}^{\lbrace i_1i_2\ldots i_S\rbrace}
&= -2 (-e)^N\int_0^\infty 
\frac{dT}{T}\,
{(4\pi T)}^{-{D\over 2}}
\,{\rm e}^{-m^2T}\prod_{i=1}^N\int_0^Td\tau_i\,
\nonumber\\& \times
W (k_{i_1},\varepsilon_{i_1};\ldots;k_{i_S},\varepsilon_{i_S}) 
P_{NS}^{\lbrace i_1i_2\ldots i_S\rbrace} 
{\rm e}^{\half \sum_{i,j=1}^N G_{ij}k_i\cdot k_j}\, .
\label{GammaNSexpl}
\end{align}
Here the polynomials 
$P_{NS}^{\lbrace i_1i_2\ldots i_S\rbrace}$ 
are now defined by (compare with \eqref{defPN})
\bear
&&{\rm e}^{\sum_{i,j=1}^N\big(-i\dot G_{ij}\varepsilon_i\cdot k_j+\half \ddot G_{ij}\varepsilon_i\cdot \varepsilon_j\big)}
\big \vert_{\varepsilon_{i_1}= \cdots = \varepsilon_{i_S} =0}\big \vert_{\varepsilon_{i_{S+1}}\cdots\varepsilon_{i_N}}
\equiv
(-i)^{N-S} P_{NS}^{\lbrace i_1i_2\ldots i_S\rbrace } \, ,
\nonumber\\
\label{defPNS}
\ear
where the notation on the left-hand side means that 
one first sets the polarisation vectors $\varepsilon_{i_1},\ldots,\varepsilon_{i_S}$ equal to zero,
and then selects all the terms linear in the surviving polarisation vectors. 
In particular, one has the extremal cases $P^{\lbrace \rbrace}_{N0} = P_N \, , P^{\lbrace 12\ldots N\rbrace}_{NN} = 1$.
For the Wick-contraction of the spin interaction terms we have introduced the notation
\bear
W(k_{i_1},\varepsilon_{i_1};\ldots;k_{i_S},\varepsilon_{i_S}) 
\equiv  \Bigl\langle 
\psi_{i_1}\cdot f_{i_1} \cdot \psi_{i_1}
\cdots
\psi_{i_S}\cdot f_{i_S} \cdot \psi_{i_S}
\Bigr\rangle
\quad\quad
\label{defW}
\ear
to be evaluated with the basic correlator 
\bear
\langle \psi^\mu(\tau) \psi^\nu(\tau') \rangle = \half G_F(\tau,\tau') \delta^\mn, \quad
G_F(\tau,\tau') \equiv {\rm sgn}(\tau-\tau') \, .
\label{wickpsi}
\ear
This object possesses the following closed-form description. 
Define a ``Lorentz cycle of length $n$'' $Z_n$ by 
\bear
Z_n(i_1i_2\ldots i_n)&\equiv&
\Bigl(\half\Bigr)^{\delta_{n2}}
{\rm tr}
\Bigl(
\prod_{j=1}^n
f_{i_j}\Bigr) 
\label{defZn}
\ear\no
and a ``fermionic bi-cycle of length $n$'' by
\bear
G_F(i_1i_2\ldots i_n) \equiv G_{Fi_1i_2}G_{Fi_2i_3}\cdots G_{Fi_ni_1}Z_n(i_1i_2\ldots i_n) \qquad (n\geq 2)\,.
\label{deffermionicbicycle}
\ear
Then we can write
\bear
W\argS 
&=& 
\sum_{\rm partitions}
(-1)^{cy} G_F(i_1i_2\ldots i_{n_1})G_F(i_{n_1+1}\ldots i_{n_1+n_2})\cdots\nonumber\\ &&\times
G_F(i_{n_1+\ldots + n_{cy-1}+1}\ldots i_{n_1+\ldots + n_{cy}})\, .
\label{Wsum}
\ear
Here the sum runs over 
all inequivalent possibilities to distribute the indices $1,\ldots,S$ among the arguments of any number $cy$ of bi-cycles,
and $n_k$ denotes the length of the bi-cycle $k$.
Working out \eqref{Wsum} up to $S=4$, we find
\bear
W (k_1,\varepsilon_1) &=& 0\, , \nonumber\\
W (k_1,\varepsilon_1;k_2,\varepsilon_2) &=& - G_F(12) \, ,  \nonumber\\
W (k_1,\varepsilon_1;k_2,\varepsilon_2;k_3,\varepsilon_3) &=& - G_F(123)\, ,   \nonumber\\
W (k_1,\varepsilon_1;k_2,\varepsilon_2;k_3,\varepsilon_3;k_4,\varepsilon_4) &=& - G_F(1234) - G_F(1243) - G_F(1324) \nonumber\\
&& + G_F(12)G_F(34) +  G_F(13)G_F(24) +  G_F(14)G_F(23)\, . \nonumber\\
\label{Weamp}
\ear

\section{$N$-photon amplitude in a plane-wave background (scalar QED)}
\label{scalar}

In general,  a plane-wave field can be defined by a vector potential $A(x)$ of the form
\bear
e A_{\mu}(x) = a_{\mu}(n\cdot x)
\ear
where $n^{\mu}$ is a null vector, 
%
\bear
n^2 = 0 
\label{nnull}
\ear
and, as is usual, we will further impose the {\it light-front gauge condition} 
\bear
n\cdot a = 0
\label{lgc}
\ear
Note that we absorb the charge $e$ in the definition of $a_\mu$. 
Repeating the procedure of the previous section with the addition of the potential $a_\mu$ to
the worldline Lagrangian, we straightforwardly get a representation of the $N$-photon amplitude in
the plane-wave background that generalizes the vacuum formula \eqref{Nphotonvertop},
\bear
\Gamma_{\rm scal}(\lbrace{k_i,\varepsilon_i\rbrace};a)
 &=&
(-ie)^N 
\int_0^{\infty}
\frac{dT}{T}\,
\e^{-m^2T}
\int
Dx\,
\e^{-\int_0^T d\tau \bigl\lbrack\kinb +i \dot x^{\mu}a_{\mu}(n\cdot x)\bigr\rbrack}
\nonumber\\
&& \times
 V^{\gamma}_{\rm scal}[k_1,\varepsilon_1]\cdots V^{\gamma}_{\rm scal}[k_N,\varepsilon_N]
\label{12-Nphotonpw}
\ear
Fixing the zero-mode problem as usual by 
separating off the average position $x_0^\mu$ of the trajectory, $x^\mu(\tau)=x_0^\mu + q^\mu(\tau)$,
we note that, differently from the vacuum case, it now appears not only in the exponents of the vertex operators, 
but also in the argument of $a_\mu (n\cdot x)$. Thus it will now be convenient to introduce 
(euclidean) light-cone coordinates adapted to the null vector $n^\mu$. 
Thus we set
$n^\mu \equiv \frac{1}{\sqrt{2}}(0,0,1,i)$, and define $x^+ \equiv n\cdot x = \frac{1}{\sqrt{2}}(x^3+ix^4)$ (``light-front time'') 
and $x^- \equiv \frac{1}{\sqrt{2}}(- x^3+ix^4)$.  
We will further denote $x^{\perp} \equiv (x^1,x^2)$. 
Defining also $k^\pm \equiv \frac{1}{\sqrt{2}} (\pm k^3 + ik^4)$,
and using the decomposition
\bear
k\cdot x = - k^+x^- - k^-x^+ + k^1x^1 + k^2x^2
\label{12-kx}
\ear
allows us to integrate out $x_0^\mu$ but for its $x_0^+$ component:
\bear
&&
\int
Dx\,
\e^{-\int_0^T d\tau \bigl\lbrack\kinb +i \dot x^{\mu}a_{\mu}(n\cdot x)\bigr\rbrack}
 V^{\gamma}_{\rm scal}[k_1,\varepsilon_1]\cdots V^{\gamma}_{\rm scal}[k_N,\varepsilon_N]
 \nonumber\\
 &&
 =
 (2\pi)^3
 \delta\bigl(\sum_{i=1}^N k_i^1\bigr)
\delta\bigl(\sum_{i=1}^N k_i^2\bigr)
\delta\bigl(\sum_{i=1}^N k_i^+\bigr)
\totint dx_0^+ 
\e^{-i x_0^+ \sum_{i=1}^N k_i^-}
\Nint
\nonumber\\&&\times
\int Dq 
\e^{-\int_0^T d\tau \bigl\lbrack\frac{\dot q^2}{4} +i \dot q^{\mu}a_{\mu}(x_0^+ + n\cdot q(\tau))\bigr\rbrack}
\,\e^{\sum_{i=1}^N (ik_i \cdot q_i + \varepsilon_i\cdot \dot q_i)}
\Big\vert_{\varepsilon_1\varepsilon_2\cdots \varepsilon_N}\,. 
\label{12-intx0}
\ear
The calculation of the functional integral at first sight looks like an intractable problem, since the integration variable $q(\tau)$
appears in the argument of the unknown function $a_\mu$. In \cite{ildtor} this problem was solved for the two-point case
using the fact that the plane-wave path integral possesses the gaussian property that its semiclassical approximation
is exact. For the $N$-point generalization, we find it more convenient to exhibit the crypto-gaussian nature of the path integral
using the relations \eqref{nnull} and \eqref{lgc}. In principle, we could do the functional integral by expanding
\bear
\e^{-\int_0^T d\tau  i \dot q\cdot a(x_0^+ + n\cdot q(\tau))}
&=&
1 - i \int_0^T d\tau \, \dot q\cdot a(x_0^+ + n\cdot q(\tau))
\nonumber\\&& \hspace{-140pt}
+
\frac{(-i)^2}{2!} \int_0^Td\tau  \, \dot q_1\cdot a(x_0^+ + n\cdot q_1)  \int_0^Td\tau'  \, \dot q_2\cdot a(x_0^+ + n\cdot q_2)
+
\cdots 
\label{12-expexpand}
\ear
and then Taylor-expanding
\bear
 a_\mu(x_0^+ + n\cdot q_m) = a_\mu(x_0^+) + a'_\mu(x_0^+)(n\cdot q_m) + \frac{1}{2!} a''_\mu(x_0^+)(n\cdot q_m)^2 + \cdots
 \label{12-taylora}
\ear 
(Note that we use a `prime' for the derivative of a function with respect to its argument, while the `dot' will be used for the total
derivative with respect to proper time.)
Now, we observe that a factor $n\cdot q_m$ can neither be Wick contracted with another such factor because of \eqref{nnull}, nor
with a factor of $\dot q_n\cdot a^{(k)}(x_0^+)$ because of \eqref{lgc}. Thus each $n\cdot q_m$ has to be contracted with the
exponential $\e^{\sum_{i=1}^N (ik_i \cdot q_i + \varepsilon_i\cdot \dot q_i)}$, and this will convert it into 
\bear
n\cdot q_m \longrightarrow n\cdot \sum_{i=1}^N\bigl[-i G_{mi}k_i + \dot G_{mi}\varepsilon_i\bigr]
\ear
We can then resum \eqref{12-taylora} into
\bear
 a_\mu(x_0^+ + n\cdot q_m) \longrightarrow a_\mu\Bigl(x_0^+ + n\cdot \sum_{i=1}^N[-i G_{mi}k_i + \dot G_{mi}\varepsilon_i]\Bigr)
 \label{changearg}
 \ear
 and subsequently also \eqref{12-expexpand},
 \bear
\e^{-\int_0^T d\tau  i \dot q\cdot a(x_0^+ + n\cdot q(\tau))}
\longrightarrow
\e^{-\int_0^T d\tau  i \dot q\cdot a 
\bigl(x_0^+  + n\cdot \sum_{i=1}^N[-i G(\tau,\tau_i)k_i + \dot G(\tau,\tau_i)\varepsilon_i]\bigr)
}
\nonumber\\
\ear
where we can now, with some abuse of notation, replace
\bear
a_\mu 
\Bigl(x_0^+  + n\cdot \sum_{i=1}^N[-i G(\tau,\tau_i)k_i + \dot G(\tau,\tau_i)\varepsilon_i]\Bigr)
\equiv 
a_\mu(\tau)
\label{12-abuse}
\ear
Thus we have removed the functional integration variable from the argument of $a_\mu$,
and converted the functional integral \eqref{12-intx0} into a gaussian one. Now the usual
``completing-the-square'' procedure can be applied, and yields
\bear
&&
\int Dq 
\e^{-\int_0^T d\tau \bigl\lbrack\frac{\dot q^2}{4} +i \dot q\cdot a(\tau)
\bigr\rbrack}
\,\e^{\sum_{i=1}^N (ik_i \cdot q_i + \varepsilon_i\cdot \dot q_i)}
\nonumber\\
&&
=
{(4\pi T)}^{-{D\over 2}}
\e^{-\frac{1}{2} \int_0^Td\tau\int_0^Td\tau' \ddot G(\tau,\tau')a(\tau)\cdot a(\tau')
-\sum_{i=1}^N \int_0^Td\tau 
\bigl\lbrack \dot G(\tau,\tau_i) a(\tau)\cdot k_i + i \ddot G(\tau,\tau_i) a(\tau)\cdot \varepsilon_i \bigr\rbrack
}
\nonumber\\
&&\quad \times
\e^{
\sum_{i,j=1}^N 
\bigl\lbrack  \half G_{ij} k_i\cdot k_j
-i\dot G_{ij}\varepsilon_i\cdot k_j
+\half\ddot G_{ij}\varepsilon_i\cdot\varepsilon_j
\bigr\rbrack}
\label{12-qintfin}
\ear
The first term in the exponent on the right-hand-side can, introducing the worldline average
\bear
\langle\langle f \rangle\rangle
\equiv
\frac{1}{T} \int_0^Td\tau f(\tau)
\label{12-wa}
\ear
and using \eqref{GdGdd} be rewritten as
\bear
\half \int_0^Td\tau\int_0^Td\tau' \ddot G(\tau,\tau')a(\tau)\cdot a(\tau')
=
T\Bigl(\langle\langle a^2 \rangle\rangle - \langle\langle a \rangle\rangle^2\Bigr)
\ear
Similarly, we can rewrite 
\bear
\sum_{i=1}^N \int_0^Td\tau \,
\ddot G(\tau,\tau_i) a(\tau)\cdot \varepsilon_i 
=
2 \sum_{i=1}^N\Bigl(a(\tau_i)-\langle\langle a\rangle\rangle \Bigr) \cdot \varepsilon_i
\ear
For the integral involving $\dot G(\tau,\tau_i)$, we introduce the periodic integral function
\bear
I_\mu(\tau) \equiv \int_0^\tau d\tau' \Bigl(a_\mu(\tau') - \langle\langle a_\mu\rangle\rangle \Bigr )
\ear
Integrating by parts, we get
\bear
\sum_{i=1}^N \int_0^Td\tau \,\dot G(\tau,\tau_i) a(\tau)\cdot k_i
=
-2
\sum_{i=1}^N k_i \cdot 
\bigl(I(\tau_i)-\langle\langle I \rangle\rangle \bigr)
\label{IBPI}
\ear
Putting the pieces together, we get the following master formula for the scalar QED
$N$-photon amplitude in a plane-wave background\footnote{The reader familiar with the worldline formalism may wonder why we could
drop the additive constant $\frac{T}{6}$ from this Green's function, which is customary but relies on
momentum conservation. It is easy to verify that here, in light-cone coordinates, the removal of the constant requires only
that $\sum_{i=1}^N k_i^{+,\perp} =0$, not full momentum conservation.}
\bear
\Gamma_{\rm scal}(\lbrace{k_i,\varepsilon_i\rbrace};a) &=&
(-ie)^N 
(2\pi)^3 
\delta\bigl(\sum_{i=1}^N k_i^1\bigr)
\delta\bigl(\sum_{i=1}^N k_i^2\bigr)
\delta\bigl(\sum_{i=1}^N k_i^+\bigr)
\totint dx_0^+ \e^{-i x_0^+ \sum_{i=1}^N k_i^-}
\nonumber\\&&\hspace{-80pt}\times
\int_0^{\infty}
\frac{dT}{T}\,
{(4\pi T)}^{-{D\over 2}}
\prod_{i=1}^N \int_0^Td\tau_i
\e^{
\sum_{i,j=1}^N 
\bigl\lbrack  \half G_{ij} k_i\cdot k_j
-i\dot G_{ij}\varepsilon_i\cdot k_j
+\half\ddot G_{ij}\varepsilon_i\cdot\varepsilon_j
\bigr\rbrack}
\nonumber\\&&\hspace{-80pt}\times
\e^{-\bigl(m^2+ \langle\langle a^2 \rangle\rangle - \langle\langle a \rangle\rangle^2\bigr)T+2\sum_{i=1}^N k_i \cdot 
\bigl(I(\tau_i)-\langle\langle I \rangle\rangle \bigr)
-2i \sum_{i=1}^N\bigl(a(\tau_i)-\langle\langle a\rangle\rangle \bigr) \cdot \varepsilon_i
}
\Bigl\vert_{\varepsilon_1\cdots \varepsilon_N}
\nonumber\\
\label{Nphotonpw}
\ear
Note that the appearance of the polarization vectors in the argument of $a_\mu$ makes it still messy to extract the terms
linear in all of them. Further substantive simplification can be achieved by choosing the $\varepsilon_i$ such as to obey
\bear
n\cdot \varepsilon_i = 0 \quad (i=1,\ldots,N)
\label{polgauge}
\ear
which is possible for generic momenta by a gauge transformation, and will be assumed for the rest of this paper. 
This will reduce \eqref{12-abuse} to
\bear
a_\mu(\tau)
=
a_\mu 
\Bigl(x_0^+  -i \sum_{i=1}^N G(\tau,\tau_i)k_i^+ \Bigr)
\label{abusesimp}
\ear
The master formula \eqref{Nphotonpw} can then be written more explicitly as
\bear
\Gamma_{\rm scal}(\lbrace{k_i,\varepsilon_i\rbrace};a)
\!\!\! &=& \!\!\! (-e)^N
(2\pi)^3 
\delta\bigl(\sum_{i=1}^N k_i^1\bigr)
\delta\bigl(\sum_{i=1}^N k_i^2\bigr)
\delta\bigl(\sum_{i=1}^N k_i^+\bigr)
\!\!
\totint dx_0^+ \e^{-i x_0^+ \sum_{i=1}^N k_i^-}
\nonumber\\
\hspace{-1em}&&\hspace{-50pt} \times
\int_0^\infty
\frac{dT}{T}\,
{(4\pi T)}^{-{D\over 2}}
\e^{-\bigl(m^2+ \langle\langle a^2 \rangle\rangle - \langle\langle a \rangle\rangle^2\bigr)T}
\prod_{i=1}^N\int_0^Td\tau_i\,
\nonumber\\
\hspace{-1em}&&\hspace{-50pt} \times
\mathfrak{P}_{N} 
\,{\rm e}^{\half \sum_{i,j=1}^N G_{ij}k_i\cdot k_j+2\sum_{i=1}^N k_i \cdot 
\bigl(I(\tau_i)-\langle\langle I \rangle\rangle \bigr)}
\label{GammaNexplpw}
\ear
where the polynomials $\mathfrak{P}_{N}$ are defined by  (compare \eqref{defPN})
\bear
{\rm e}^{\sum_{i,j=1}^N\big(-i\dot G_{ij}\varepsilon_i\cdot k_j+\half \ddot G_{ij}\varepsilon_i\cdot \varepsilon_j\big)
-2i \sum_{i=1}^N\bigl(a(\tau_i)-\langle\langle a\rangle\rangle \bigr) \cdot \varepsilon_i
}
\big \vert_{\varepsilon_{1}\cdots\varepsilon_{N}}
\equiv
(-i)^{N} {\mathfrak{P}}_{N} \, .
\label{defPNpw}
\ear

\section{$N$-photon amplitude in a plane-wave background (spinor QED)}
\label{spinor}

Proceeding to the spinor QED case, the same argument that we applied above to $eA_\mu =a_\mu$
can be used to convert also the argument of the $eF_\mn = n_\mu a'_\nu - a'_\mu n_\nu$ appearing in the
spin part of the worldline Lagrangian in \eqref{spinorQEDgrass} in the same way as in \eqref{changearg}. 
Thus the only new element is that the fermionic Wick-contraction rule \eqref{defW} now has to be calculated with a
generalised worldline Green's function inverting the field-dependent operator 
\bear
{\cal O} \equiv  \frac{\delta_\mn}{2}  \ddtau + i a'_\mu(\tau)n_\nu  - i n_\mu a'_\nu (\tau)
\label{defO}
\ear
The appropriate generalisation of \eqref{wickpsi} is 
\bear
\langle \psi^\mu(\tau) \psi^\nu(\tau') \rangle = \half \mathfrak G_F^\mn(\tau,\tau'), 
\label{wickpsigen}
\ear
where
\begin{equation}
\hspace{-1.5em}\mathfrak G_F^\mn(\tau,\tau') 
\equiv 
\biggl\lbrace \delta^\mn + 2i n^\mu{\cal J}^\nu(\tau,\tau') + 2i {\cal J}^\mu(\tau',\tau)n^\nu
+ 2\Bigl\lbrack {\cal J}^2(\tau,\tau')-\frac{T^2}{4} \langle\langle a'\rangle\rangle^2\Bigr\rbrack  
n^\mu n^\nu\biggr\rbrace 
G_F(\tau,\tau')
\label{Gfplane}
\end{equation}
and we have further defined
\bear
J_\mu(\tau) &\equiv& \int_0^\tau d\tau' \Bigl( a'_\mu(\tau') - \langle\langle a'_\mu \rangle\rangle \Bigr)\, , \\
{\cal J}_\mu(\tau,\tau') &\equiv& J_\mu(\tau)-J_\mu(\tau') - \frac{T}{2}\dot G (\tau,\tau')  \langle\langle a'_\mu \rangle\rangle \, .
\ear
Note that the modified Green's function has a non-zero coincidence limit,
\begin{equation}
	\mathfrak{G}_F^\mn(\tau,\tau) = -iT\Big(n^{\mu}\langle\langle a'^{\nu} \rangle\rangle - \langle\langle a'^\mu \rangle\rangle n^{\nu}\Big)
\end{equation}
and satisfies the anti-symmetry relation $\mathfrak{G}_F(\tau',\tau)  = -\mathfrak{G}_F^{\intercal}(\tau,\tau')$. 
Proceeding as in the vacuum case, we get a spin-orbit decomposition as in \eqref{decompGammaSO} with
\bear
\Gamma_{NS}^{\lbrace i_1i_2\ldots i_S\rbrace}
&=& -2 (-e)^N
(2\pi)^3 
\delta\bigl(\sum_{i=1}^N k_i^1\bigr)
\delta\bigl(\sum_{i=1}^N k_i^2\bigr)
\delta\bigl(\sum_{i=1}^N k_i^+\bigr)
\totint dx_0^+ \e^{-i x_0^+ \sum_{i=1}^N k_i^-}
\nonumber\\&&\hspace{-50pt} \times
\int_0^\infty
\frac{dT}{T}\,
{(4\pi T)}^{-{D\over 2}}
\e^{-\bigl(m^2+ \langle\langle a^2 \rangle\rangle - \langle\langle a \rangle\rangle^2\bigr)T}
\prod_{i=1}^N\int_0^Td\tau_i\,
\nonumber\\&&\hspace{-50pt} \times
\mathfrak{W} (k_{i_1},\varepsilon_{i_1};\ldots;k_{i_S},\varepsilon_{i_S}) 
\mathfrak{P}_{NS}^{\lbrace i_1i_2\ldots i_S\rbrace} 
{\rm e}^{\half \sum_{i,j=1}^N G_{ij}k_i\cdot k_j+2\sum_{i=1}^N k_i \cdot 
\bigl(I(\tau_i)-\langle\langle I \rangle\rangle \bigr)}
\, .
\label{GammaNSexplpw}
\ear
The polynomials $\mathfrak{P}_{NS}$ now are defined by 
\bear
\hspace{-1.5em}{\rm e}^{\sum_{i,j=1}^N\big(-i\dot G_{ij}\varepsilon_i\cdot k_j+\half \ddot G_{ij}\varepsilon_i\cdot \varepsilon_j\big)
-2i \sum_{i=1}^N\bigl(a(\tau_i)-\langle\langle a\rangle\rangle \bigr) \cdot \varepsilon_i
}
\big \vert_{\varepsilon_{i_1}= \cdots = \varepsilon_{i_S} =0}\big \vert_{\varepsilon_{i_{S+1}}\cdots\varepsilon_{i_N}}
\equiv
(-i)^{N-S} {\mathfrak{P}}_{NS}^{\lbrace i_1i_2\ldots i_S\rbrace } \, ,
\nonumber\\
\label{defPNSpw}
\ear
and $\mathfrak{W} (k_{i_1},\varepsilon_{i_1};\ldots;k_{i_S},\varepsilon_{i_S})$ denotes the correlator \eqref{defW} evaluated
with the modified fermionic Wick contraction \eqref{wickpsigen}. For the calculation of this correlator we can
still use the cycle decomposition formula \eqref{Wsum}, only that the fermionic bicycle 
\eqref{deffermionicbicycle}, now must be replaced by
\begin{eqnarray}
\mathfrak{G}_F(i_1i_2\dots i_{n})&~\equiv~ \bigl(\half\bigr)^{\delta_{n2}}
\textrm{tr}(f_{i_1}\cdot{\mathfrak{G}}_{Fi_1i_2}\cdot f_{i_2}\cdot{\mathfrak{G}}_{Fi_2i_3}\cdots f_{i_n}\cdot{\mathfrak{G}}_{Fi_ni_1})~~
\label{genbicycle}
\end{eqnarray}
Note that, differently from the case of a constant external field \cite{shaisultanov,18}, the fermionic path-integral determinant factor is not affected by the presence of the plane-wave field, and remains at its free value $2^\frac{D}{2}$.

\section{The case $N=2$}
\label{N2}

As a check, let us show that the above master formulas correctly reproduce the results of \cite{ildtor} for the $N=2$ case. We first give the general off-shell results before specialising to the on-shell helicity flip process studied there. We work throughout with the gauge choice \eqref{polgauge} for convenience and note that momentum conservation in the $+$ direction gives $k_{2}^{+} = -k_{1}^{+}$. 

\subsection{Off-shell}

Using the notation introduced in \eqref{decompGammaSO}, for $\Gamma_{20}$ we find from \eqref{defPNSpw}
\begin{align}
	\mathfrak{P}_{20}^{\{\}} &= \dot{G}_{12}\dot{G}_{21}\varepsilon_{1}\cdot k_{2} \varepsilon_{2}\cdot k_{1} - \ddot{G}_{12}\varepsilon_{1}\cdot \varepsilon_{2}\nonumber \\
	&+2 \Big(\dot{G}_{12}\varepsilon_{1}\cdot k_{2} \varepsilon_{2}\cdot \big(a(\tau_{2}) - \langle\langle a\rangle \rangle \big) + \dot{G}_{21}\varepsilon_{1}\cdot \big(a(\tau_{1}) - \langle \langle a \rangle \rangle \big) \varepsilon_{2}\cdot k_{1}\Big) \nonumber \\
	&+4 \varepsilon_{1}\cdot \big(a(\tau_{1}) - \langle \langle a \rangle \rangle \big)\varepsilon_{2}\cdot \big(a(\tau_{2}) - \langle\langle a\rangle \rangle \big)\\
	\mathfrak{W}() &= 1\,,
\end{align}
which is sufficient to 
produce the scalar QED result when substituted into (\ref{GammaNSexpl}). Furthermore, for the spinor case we also require
\begin{align}
	\mathfrak{P}_{21}^{\{1\}} &= \dot{G}_{21}\varepsilon_{2}\cdot k_{1} + 2\varepsilon_{2}\cdot \big(a(\tau_{2}) - \langle\langle a\rangle \rangle \big)\\
	\mathfrak{W}(k_{1}, \varepsilon_{1}) &= -\frac{1}{2} \textrm{tr}\big(\mathfrak{G}_{F}(\tau_{1}, \tau_{1})\cdot f_{1}\big) 
	= -iTk_{1}^{+}\varepsilon_{1}\cdot \langle \langle a' \rangle \rangle
\end{align}
which, with the corresponding results under $1 \leftrightarrow 2$, can be used to construct $\Gamma_{21}$. Finally, we determine $\Gamma_{22}$ from
\begin{align}
	\mathfrak{P}_{22}^{\{12\}} &= 1\\
	\mathfrak{W}(k_{1}, \varepsilon_{1}; k_{2}, \varepsilon_{2}) &= - \mathfrak{G}_{F}(12) + \frac{1}{4}\textrm{tr}\big(\mathfrak{G}_{F}(\tau_{1}, \tau_{1}) \cdot f_{1}\big) \textrm{tr}\big(\mathfrak{G}_{F}(\tau_{2}, \tau_{2})\cdot f_{2}\big)\,
\end{align}
where ($\mathcal{J}_{ij} \equiv \mathcal{J}(\tau_{i}, \tau_{j})$)
\begin{align}
	-\mathfrak{G}_{F}(12) &=  2k_{1}^{+}k_{2}^{+} \Big[\varepsilon_{1} \cdot \varepsilon_{2} \Big(\mathcal{J}_{12}^{2} - \frac{T^{2}}{4} \langle \langle a'\rangle \rangle^{2}\Big) + 2 \varepsilon_{1}\cdot \mathcal{J}_{21} \varepsilon_{2}\cdot \mathcal{J}_{12}\Big] \sigma_{12}\sigma_{21}               \nonumber \\
	 &- 2i k_{1}^{+}\Big[\varepsilon_{1}\cdot k_{2} \varepsilon_{2}\cdot \mathcal{J}_{12} - \varepsilon_{1}\cdot \varepsilon_{2} k_{2}\cdot \mathcal{J}_{12}\Big]  \sigma_{12}\sigma_{21}              \nonumber \\
	&- 2ik_{2}^{+}\Big[\varepsilon_{2}\cdot k_{1}\varepsilon_{1}\cdot \mathcal{J}_{21} - \varepsilon_{1}\cdot \varepsilon_{2} k_{1}\cdot \mathcal{J}_{21}\Big] \sigma_{12}\sigma_{21}              \nonumber \\
	&-\frac{1}{2}\textrm{tr}\big(f_{1}\cdot f_{2}\big) \sigma_{12}\sigma_{21}   \,.
	\label{W2background}
\end{align}
These lead to a spin-orbit decomposition
\begin{align}
\Gamma_{\textrm{spin}, 2}(k_{1}, \varepsilon_{1}; k_{2}, \varepsilon_{2}) &= -2 e^2 \int_{-\infty}^{\infty}dx_{0}^{+}\, \e^{-ix_{0}^{+} (k_{1} + k_{2})^{-}} \int_0^\infty \frac{dT}{T} (4\pi T)^{-\frac{D}{2}}
\nonumber\\
& \hspace{-80pt} \times
 {\rm e}^{-\big(m^2 + \langle \langle a^{2} \rangle \rangle - \langle \langle a\rangle \rangle^2 \big) T}
\int_0^Td\tau_1\int_0^Td\tau_2 
\, {\rm e}^{G_{12}k_{1}\cdot k_{2}+2\sum_{i=1}^2 k_i \cdot 
\bigl(I(\tau_i)-\langle\langle I \rangle\rangle \bigr)}
\nonumber\\
&\hspace{-80pt} \times
\Big\{ \mathfrak{P}_{20}^{\{\}} + \mathfrak{W}(k_{1}, \varepsilon_{1})\mathfrak{P}_{21}^{\{1\}} + \mathfrak{W}(k_{2}, \varepsilon_{2})\mathfrak{P}_{21}^{\{2\}} + \mathfrak{W}(k_{1}, \varepsilon_{1}; k_{2}\varepsilon_{2})\Big\}\,, 
\label{soN2}
\end{align}
where we have omitted the momentum conserving $\delta$-functions in the $+$ and $\perp$ directions.  
The term in the exponent 
$\sum_{i=1}^{2}k_{i}\cdot \big(I(\tau_{i}) -\langle \langle I \rangle \rangle \big) $ 
at the two-point level could be removed by imposing on $a_\mu$, instead of \eqref{lgc}, the stronger gauge condition of full transversality.  
To see this, it is easiest to return to \eqref{IBPI}. Using the conservation of momentum along the
$+$ and transversal directions together with \eqref{abusesimp}, and choosing a function $b_\mu(x)$ such that $b'_\mu=a_\mu$, we have 
\bear  
\sum_{i=1}^2 \int_0^Td\tau \,\dot G(\tau,\tau_i) a(\tau)\cdot k_i
&=&
\int_0^Td\tau \,(\dot G(\tau,\tau_1)-\dot G(\tau,\tau_2)) a(\tau)\cdot k_1
\nonumber\\
&=&
\frac{i}{k_1^+}
\int_0^Td\tau \frac{d}{d\tau} k_1\cdot b \Bigl(x_0^+  -i \sum_{i=1}^N G(\tau,\tau_i)k_i^+ \Bigr)
\nonumber\\
&=& \frac{i}{k_1^+} k_1\cdot (b(T) - b(0)) =
0
\ear

\subsection{On-shell}
In the on-shell case and for $N=2$ photons we gain additional simplifications due to the mass shell condition which, by conservation of momentum in the $+$ and $\perp$ directions, implies the additional condition $k_{2}^{-} = -k_{1}^{-}$ so that $k_{1} = k = -k_{2}$ with $k^{2} = 0$. This removes the exponent $\e^{G_{12}k_1\cdot k_2}$ from (\ref{soN2}). Further imposing the transversality conditions $\varepsilon_{1}\cdot k = 0 = k\cdot \varepsilon_{2}$, the components of the spin orbit decomposition reduce to
\begin{align}
	\hspace{-1em}\mathfrak{P}_{20}^{\{\}} &\rightarrow -\ddot{G}_{12}\varepsilon_{1}\cdot \varepsilon_{2} + 4\varepsilon_{1}\cdot \big(a_{1} - \langle\langle a \rangle \rangle \big) \varepsilon_{2}\cdot \big(a_{2} - \langle \langle a \rangle \rangle \big) \\
\hspace{-1em}	\mathfrak{P}_{21}^{\{1\}} &\rightarrow 2\varepsilon_{2}\cdot \big(a_{2} - \langle \langle a \rangle \rangle \big)
	, \quad \mathfrak{P}_{21}^{\{2\}} \rightarrow 2\varepsilon_{1}\cdot \big(a_{1} - \langle \langle a \rangle \rangle \big) \\
\hspace{-1em}	\mathfrak{W}(k_{i}, \varepsilon_{i}) &\rightarrow -iTk^{+}\varepsilon_{i}\cdot \langle \langle a\rangle \rangle \\
\hspace{-1em}	\mathfrak{W}(k_{1}, \varepsilon_{1}; k_{2}\varepsilon_{2}) &\rightarrow 
	k^{+2}\Bigl[2\varepsilon_{1}\cdot \varepsilon_{2}\big(\mathcal{J}_{12}^{2} - \frac{T^{2}}{4}\langle \langle a' \rangle \rangle^{2} \big)
	+4\varepsilon_{1}\cdot \mathcal{J}_{12}\varepsilon_{2}\cdot \mathcal{J}_{21} + T^{2}\varepsilon_1 \cdot \langle \langle a'\rangle \rangle \varepsilon_{2}\cdot \langle \langle a'\rangle \rangle \Big]
\end{align}

\subsection{Helicity flip}

For the purpose of comparing with the helicity-flip calculation of \cite{ildtor} we can further put $\varepsilon_{1} \cdot \varepsilon_{2} = 0$, resulting in
\begin{align}
\Gamma_{\textrm{spin}, 2}(k_{1}, \varepsilon_{1}; k_{2}, \varepsilon_{2}) &= -2 e^2 \int_{-\infty}^{\infty}dx_{0}^{+}
\int_0^\infty \frac{dT}{T} (4\pi T)^{-\frac{D}{2}}
\nonumber\\
& \hspace{-3em} \times
 {\rm e}^{-\big(m^2 +\langle \langle a^{2} \rangle \rangle - \langle \langle a\rangle \rangle^2 \big) T}
\Big\{ 4\varepsilon_{1}\cdot \big(a_{1} - \langle\langle a \rangle \rangle \big) \varepsilon_{2}\cdot \big(a_{2} - \langle \langle a \rangle \rangle \big)\nonumber \\
&\hspace{-3em} - 2 iTk^{+} \Big[\varepsilon_{1}\cdot \langle \langle a\rangle \rangle \varepsilon_{2}\cdot \big(a_{2} - \langle \langle a \rangle \rangle \big)-\varepsilon_{1}\cdot \big(a_{1} - \langle \langle a \rangle \rangle \big)\varepsilon_{2}\cdot \langle \langle a\rangle \rangle \Big]\nonumber \\
&\hspace{-3em} +  k^{+2} \Big[4\varepsilon_{1}\cdot \mathcal{J}_{12}\varepsilon_{2}\cdot \mathcal{J}_{21} + T^{2}\varepsilon_1 \cdot \langle \langle a'\rangle \rangle \varepsilon_{2}\cdot \langle \langle a'\rangle \rangle \Big]  \Big\}\,.
\label{soN2flip}
\end{align}
This indeed correctly reproduces the parameter integrals determined in \cite{ildtor} as can be easily seen using the integral representations
\begin{align}
	\mathcal{J}_{12 \mu} &= -\frac{1}{2}\sigma_{12}\int_{0}^{T}d\tau\, \sigma(\tau - \tau_{1})\sigma(\tau-\tau_{2}) a'_{\mu}(\tau)\nonumber \\
	\mathcal{J}_{12}^{2} - \frac{T^{2}}{4}\langle \langle a'\rangle \rangle^{2} &= -\frac{1}{2}\sigma_{12}\int_{0}^{T}d\tau \int_{0}^{T}d\tau' \, \sigma(\tau - \tau_{1})\sigma(\tau - \tau')\sigma(\tau' - \tau_{2}) a'(\tau)\cdot a'(\tau')\,.
	\label{ints}
\end{align}

\section{Summary and Outlook}

We have used the worldline formalism to construct a master formula for the $N$-photon amplitudes in a general
plane-wave background field, for both scalar and spinor QED. 
We believe that it compares favourably with other available techniques for calculations in plane-wave backgrounds.
As usual in applications of the worldline formalism to QED, it unifies the scalar and spinor cases in the sense that
any spinor QED calculation yields the corresponding scalar QED quantity as a side result. 
In particular, with the formalism developed here a calculation of the photon-photon scattering amplitudes in 
a plane-wave background should be quite feasible. 
In a more extensive publication, we will give a more detailed derivation, including an alternative approach using the
worldline super formalism for the spinor QED case, and explore the $N=3$ and $N=4$ cases. 
It should also be interesting to generalize the mapping of worldline averages to spacetime averages, introduced in \cite{ildtor}, to the
$N$-point case. 
Also under consideration is the extension of the formalism to the open-line case, i.e. the photon-dressed scalar and spinor propagators
in a plane-wave background. 

\bigskip

\no
{\bf Acknowledgements:} We thank Anton Ilderton for helpful correspondence concerning some of the details in \cite{ildtor}.





\end{document}